\documentclass[%
 reprint,
 amsmath,amssymb,
 aps,
beamer,
]{revtex4-1}

\usepackage{setspace}
\usepackage{graphicx}
\usepackage{latexsym}
\usepackage{color}
\usepackage[table]{xcolor}
\usepackage{amsmath,amsthm,amssymb}
\usepackage{accents}
\usepackage{flafter}
\usepackage{amssymb}
\usepackage{bbold} 
\usepackage{bm}
\usepackage{pstricks}
\usepackage{pst-plot}
\usepackage{cancel}
\usepackage{eufrak,mathrsfs,mathrsfs}
\usepackage{dsfont}    
\usepackage{makeidx}         
\usepackage{nomencl}         
\usepackage{float}
\usepackage{xfrac}
\usepackage{hyperref}
\usepackage{xcolor}
\definecolor{red}{rgb}{1.0,0.0,0.0}
\definecolor{blue}{rgb}{0.0,0.0,1.0}
\definecolor{dark-gree}{rgb}{0.0,0.5,0.0}
\hypersetup{
    colorlinks, 
    linkcolor={red},
    citecolor={blue}, 
    urlcolor={dark-gree}
}

\usepackage{graphicx}
\usepackage{dcolumn}
\usepackage{bm}

\newcommand{\lb}[1]{\label{#1}}

\newcommand{\bra}{\langle}
\newcommand{\ket}{\rangle}

\begin{document}

\preprint{APS/123-QED}

\title{ Elucidating Dicke Superradiance by quantum uncertainty}

\author{Eduardo M. dos Santos}
\author{Eduardo I. Duzzioni}
 \email{eduardo.duzzioni@ufsc.br}
\affiliation{ Departamento de F\'{i}sica, Universidade Federal de Santa Catarina, CEP 88040-900, Florian\'{o}polis, SC, Brazil}

\date{\today}

\begin{abstract}
Recently it was shown in Ref. [\emph{Phys. Rev. Lett.} \textbf{112}, 140402 (2014)] that in the idealized Dicke model of superradiance there is no entanglement among any partitions of the system during the total evolution time of the system. This result immediately conducts us to question if other measures from quantum information theory can explain the characteristic release of energy in a short time interval. In this work we identify the uncertainty of purely quantum origin as the property responsible for Dicke superradiance. The quantum uncertainty on the population of each emitter of the sample captured by the Wigner-Yanase skew information (WYSI) is proportional to the correlation radiation rate, which is the part of the total radiated power coming from dipole correlations and responsible for releasing in a short time a great intensity of radiation energy. We also show that the correlation measure called local quantum uncertainty, which is the minimization of the WYSI over all local observables, presents a double sudden change induced by environment. The time window between these two sudden changes is used to define the interval in which symmetric global observables of the system behave classically for $N \rightarrow \infty$, although the emitters remain strongly quantum correlated.

\end{abstract}


\maketitle

Unveiling the resources behind several intriguing quantum phenomena is a hard task. It is known that for pure state quantum computation entanglement is necessary to obtain an exponential speed-up in the processing time of some quantum algorithms \cite{Shor,Feynman,Hallgren,Freedman,Aharonov,Harrow} when compared to their best classical performance \cite{Jozsa}. However, for mixed state quantum computation, necessary conditions are elusive, although quantum discord \cite{Ollivier,Henderson} could be a good candidate in the deterministic quantum computation model \cite{Datta}. There are other protocols in which discordant states are crucial to take advantage from the quantum world: the work extraction by Maxwell's demon \cite{Zurek}, entanglement activation \cite{Piani2011}, quantum metrology \cite{Modi2011}, superdense coding \cite{Meznaric}, communication cost in the quantum state merging protocol \cite{Madhok}, and quantum data hiding \cite{Piani2014}. What is interesting in all these protocols is that entanglement, which has been in the spotlight for decades, gives way to new kinds of quantum correlations \cite{Modi,Celeri}.

By its turn, Dicke superradiance is a cooperative phenomenon in which a sample composed by $N$ identical emitters can release an amount of radiation energy in a time window $N$ times shorter than the characteristic emission time of one isolated emitter and the intensity of the radiation scales with $N^2$ \cite{Dicke}. The great interest in Dicke superradiance is not only from the point of view of foundational aspects of quantum theory, but also the possibility of building superradiant lasers \cite{Bohnet} is relevant for precision measurement science. Furthermore, for efficient
collection and transfer of solar energy in light harvesting complexes, plants benefit from the superradiant processes \cite{Monshouwer,Zhao,Celardo}.

 In the particular case of superradiance in which all emitters are initially excited, the state of the system remains a linear convex combination of Dicke states the whole time. As a Dicke state (see Eq. (\ref{eq:DS})) is a highly entangled state, except the states $|J,J\ket$ and $|J,-J\ket$, it is quite natural to attribute to entanglement the power of speeding up the radiation rate. Contrary to this idea, in Ref. \cite{Wolfe} the authors show that there is no entanglement generated throughout the dynamics. Thus, the natural question is \emph{What is the property responsible for the speed up of radiation in Dicke superradiance?} We found that the Wigner-Yanase skew information (WYSI) \cite{Wignera, Araki, Wignerb} evaluated for local and global observables is a powerful tool from quantum information theory that captures the dynamical characteristics of Dicke superradiance and sheds light on the subject. As the main result of our work, we analytically show that the time dependence of the correlation radiation rate, i.e., the part of the radiated power coming from dipole correlations, is proportional to the WYSI evaluated for the local observable accounting for the population of each emitter. On the other hand, by using global observables, we show that the WYSI for the macroscopic polarization is able to signal the time of maximum emission of radiation in the classical limit. We mean by classical behavior of some observables, when the expected value of these observables can be described by a classical probability distribution in the limit $N \rightarrow \infty$ \cite{OBS}.  Further, analyzing the quantum correlations between one emitter and the rest of the system, we find a double sudden change induced by environment of the local quantum uncertainty, so that the time window between these two sudden changes can be used to define the interval in which global symmetric observables of the system behave classically. 
 
 Let us first introduce the mathematical tools and the physical system to be analyzed.\\

\emph{Wigner-Yanase Skew Information} -- With the idea of constructing a measure of information which an ensemble contains with respect to nonideal measurements, Wigner, Yanase, and Araki  introduced a measure of information $I(\rho,K)$ defined by 
\begin{equation}\lb{eq:WI}
I(\rho,K)=-\frac{1}{2}Tr\left( \left[  \sqrt{\rho},K \right]^{2} \right),
\end{equation}
where $\rho$ is the density matrix of the system and $K$ is an observable \cite{Wignera, Araki, Wignerb}. The so called \emph{Wigner-Yanase skew information} (WYSI) $I(\rho,K)$ was derived in order to be always nonnegative and to satisfy $0 \leq I(\rho,K) \leq Var(\rho,K)$, where $I(\rho,K)=0$ is achieved when $\rho$ and $K$ commute, and $I(\rho,K) = Var(\rho,K)$ when $\rho$ is a pure state, with the variance defined by $Var(\rho,K) \equiv Tr (\rho K^{2}) - \left[ Tr ( \rho K ) \right]^2$. Another important property of $I(\rho,K)$ is its convexity under mixing, i.e., $I\left( \sum_{i} p_{i}\rho_{i}, K \right) \le \sum_{i} p_{i} I\left( \rho_{i}, K \right)$, with $\sum_{i} p_{i}=1$. This means that it does not increase under classical mixing. Notice that when $\rho$ is a linear convex combination of eigenstates of $K$, $I(\rho,K)=0$, which means that for the WYSI to be non null, the density matrix must have coherence in the basis defined by $K$'s eigenstates. Therefore, the uncertainty on the observable $K$ captured by the WYSI is of quantum origin only and is an effect of quantum coherence. 

We remark that minimizing the WYSI over all local observables of a given partition, we obtain the local quantum uncertainty (LQU), a discord-like quantum correlation measure, which means that if the quantum uncertainty over all local $K$ is no-null, the subsystem is quantum correlated to the remaining system \cite{Girolamia}. The LQU presents the advantage of being analytically computable for a qubit-qudit system, which for the $l$-th two-level emitter is 
\begin{equation} \lb{eq:lqu}
\emph{LQU}_{l}(\rho) = 1 - \lambda_{max}\{ W^{l}\},
\end{equation}
where $\lambda_{max}\{ W^{l}\}$ means the maximum eigenvalue of the symmetric matrix whose entries are 
\begin{equation}
W^{l}_{\alpha \beta}=Tr\left( \sqrt{\rho}\sigma^{l}_{\alpha}  \sqrt{\rho}\sigma^{l}_{\beta} \right)
\end{equation}
with the indexes $\alpha, \beta = x,y,z$ \cite{Girolamia}. 

\emph{Dicke Superradiance}-- Let us consider the superradiant sample as composed by $N$ identical two-level emitters, where $\omega$ is the transition frequency between the ground $|g\ket$ and excited $|e\ket$ states. The operators $\sigma_{x}^{l}$, $\sigma_{y}^{l}$, and $\sigma_{z}^{l}$ are the Pauli spin matrices of the $l$-th subsystem, so that the population of $l$-th emitter is the expectation value of $\sigma_{z}^{l}$, and the ladder operators $\sigma_{\pm}^{l} \equiv \sigma_{x}^{l} \pm i\sigma_{y}^{l} $ are used to define the electric dipole operator \cite{Gross,Agarwal}. The collective operators $J_{-}$, $J_{+}$, and $J_{z}$  are defined by ($\hbar=1$)
\begin{equation} \lb{eq:Operators}
J_{\pm} \equiv \sum_{l=1}^{N} \sigma_{\pm}^{l}  \hspace{0.5cm} \text{and} \hspace{0.5cm} J_{z} \equiv \frac{1}{2} \sum_{l=1}^{N} \sigma_{z}^{l},
\end{equation}
where the Dicke states $|J,M\ket$ are eigenstates of $J_{z}$ with eigenvalues $M$ and have the form
\begin{equation} \lb{eq:DS}
|J,M\ket = \sqrt{\frac{(J+M)!}{(2J)!(J-M)!}}\left( \sum_{l=1}^{N} \sigma_{-}^{l} \right)^{(J-M)}|e,e, ...,e\ket,
\end{equation}
with $M=-J,-J+1,...,J-1,J$ and $J=N/2$. If initially all atoms are in the totally excited state, the state of the system at time  $t$ is 
\begin{equation} \lb{eq:rho}
\rho(t)=\sum_{M=-J}^{J}p_{J,M}(t)|J,M\ket \bra J,M|,
\end{equation}
where $p_{J,M}(t)$ are occupations of the Dicke states. An analytical expression for the coefficients $p_{J,M}(t)$ can be found in chapter 13 of Ref. \cite{Agarwal}.

As stated before, one signature of the superradiant emission is the total radiated power
\begin{equation} \lb{eq:ptotal}
P(t)=-\omega \frac{ d\bra J_{z} \ket}{dt} = 2\gamma \omega \bra J^{+}J^{-} \ket,
\end{equation}
which can be decomposed as the sum of two terms \cite{Delanty}. The first one
\begin{equation} \lb{eq:pind}
P_{ind} (t)= 2\gamma \omega \sum_{l=1}^{N} \bra \sigma_{+}^{l}\sigma_{-}^{l}   \ket = 2\gamma \omega \left( J + \bra J_{z} \ket \right),
\end{equation}
is the \emph{independent radiation rate}, which accounts for the radiated power by each subsystem independently of the others. The coefficient $\gamma$ is the spontaneous decay rate of each individual emitter. The second term 
\begin{equation} \lb{eq:pcorr}
P_{corr} (t)= 2\gamma \omega \sum_{l=1, l\neq k}^{N} \bra \sigma_{+}^{l}\sigma_{-}^{k}\ket = 2\gamma \omega \left( J^{2} - {\bra J_{z}^{2} \ket} \right),
\end{equation}
called \emph{correlation radiation rate}, accounts for the radiation rate emitted by pairs of correlated dipoles, which shows a quadratic dependence on $N$ \cite{Delanty}.  Indeed, this is the term responsible for releasing in a short time a great intensity of radiation energy in the superradiance phenomenon.

\emph{Results} -- Now we are able to evaluate the WYSI for the observable $\sigma_{z}$ of the $l$-th emitter. Through Eqs. (\ref{eq:WI}) and (\ref{eq:rho}), we obtain
\begin{multline}
I(\rho(t),\sigma_{z}^{l}\otimes \textbf{1}_{N-l})= \bra \left( \sigma_{z}^{l} \right)^{2} \ket \\
- \sum_{M,M^{'}=-J}^{J} \sqrt{p_{J,M}p_{J,M^{'}}} | \bra J,M | \sigma_{z}^{l} |J,M^{'}\ket|^{2}. 
\end{multline}
Using the relation $J \bra J,M | \sigma_{z}^{l}\otimes \textbf{1}_{N-l} | J,M' \ket =  \bra J,M | J_{z} | J,M' \ket = M \delta_{M,M'}$, which takes into account the indistinguishability of the emitters, we find that
\begin{equation} \lb{eq:WYlocal}
I(\rho(t),\sigma_{z}^{l}\otimes \textbf{1}_{N-l})= \frac{J^{2} - \bra J_{z}^{2} \ket}{J^{2}}.
\end{equation}
Finally, by comparing Eqs. (\ref{eq:pcorr}) and (\ref{eq:WYlocal}) we obtain our main result
\begin{equation}\lb{eq:PWYlocal}
P_{corr}(t) = \frac{\gamma \omega N^{2}}{2} I(\rho(t),\sigma_{z}^{l}\otimes \textbf{1}_{N-l}).
\end{equation}
This equation shows that the time dependence of the radiated power in the Dicke model is explained by the quantum uncertainty in the population of each emitter with respect to the whole system. In other words, the superradiant emission occurs due to the indiscernibility of the emitter with respect to photon emission. This result coming from quantum information theory perfectly agrees with the previous interpretation about the origin of the superradiance in small samples \cite{Gross}.  In FIG. 1 we observe that the WYSI for $\rho(t)$ and $\sigma_{z}^{l}$ achieves its maximum value when the radiated power is maximum (blue dotted line). The original argument given by Dicke, in which the superradiance is due to coherence, is easily seen in Eq. (\ref{eq:pcorr}), where the elements of $\bra J,M| \sigma_{+}^{l}\sigma_{-}^{k} |J,M \ket$ are non null provided that the Dicke states are coherent superpositions in the basis $\{ |g\ket, |e\ket \}^{\otimes N}$, except for $M=\pm J$. However, such coherence plays an important role only locally, which is captured by the WYSI for local observables. As will be seen later, such coherence is not crucial for global observables, since the system behaves classically around the time of maximum emission of radiation. 
\begin{figure}[h]
     \includegraphics[scale=0.43]{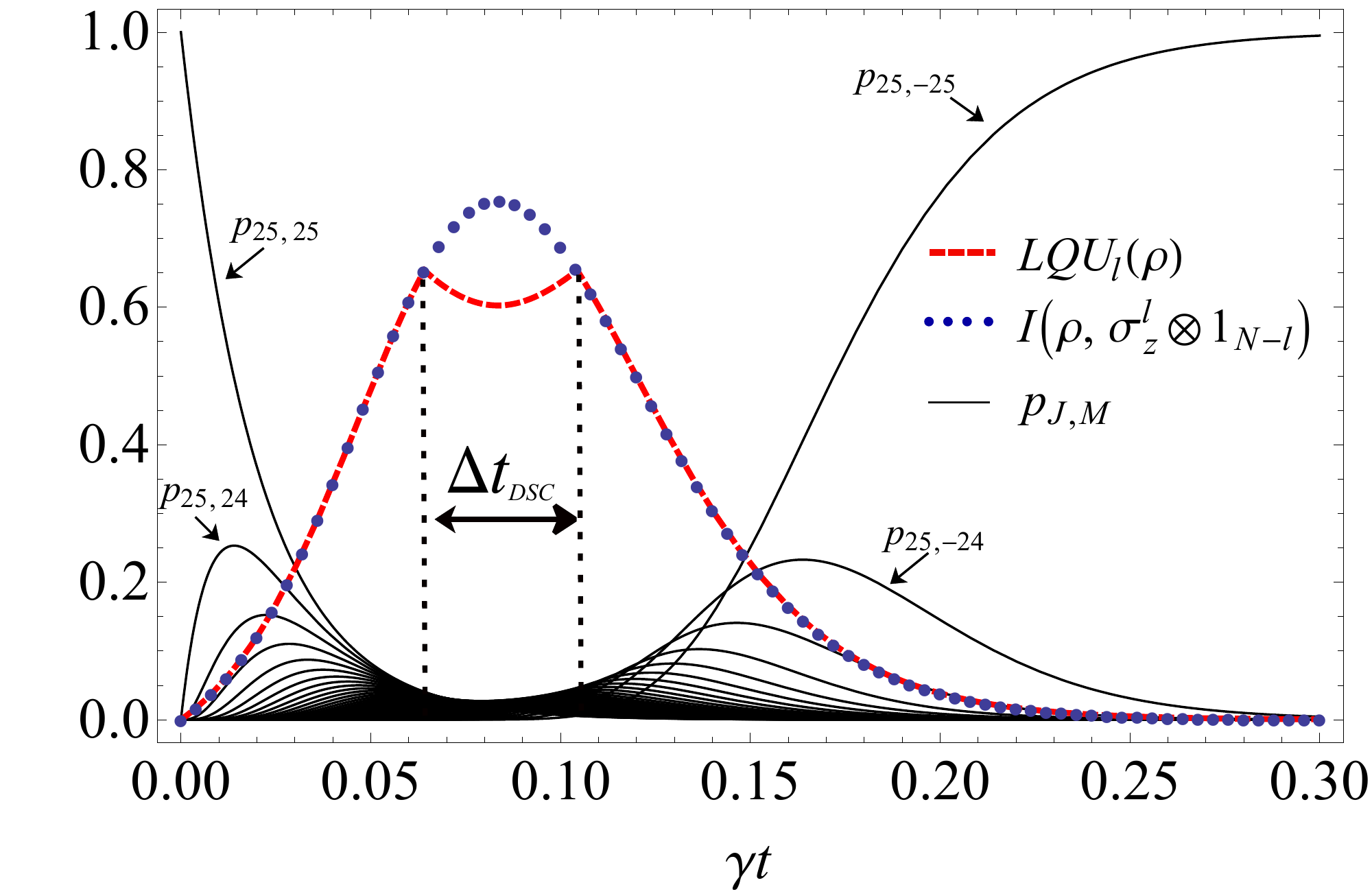}
     \label{WYJx}
 \caption{\small (Color online) The WYSI of the observable $\sigma_{z}^{l}$ calculated with $\rho(t)$ (dotted blue line), the LQU (dashed red line), and the time evolution of the populations $p_{J,M}$ of the Dicke states (black solid lines) for $N=50$ emitters in the sample. The time interval $\Delta t_{DSC}$ is a region where the symmetric global observables of the system behave classically.}
\end{figure}

Until now, the double sudden change have been observed theoretically in the original version of quantum discord \cite{Ollivier,Henderson}, as first reported in Ref.\cite{Fanchini}, and in the geometric version of quantum discord \cite{Montealegre}. Subsequently, an experimental measurement of its geometric version was reported in Ref. \cite{Paula}. In FIG. 1 we plot the LQU (red dashed line), which states that there are discord-like quantum correlations in the system, although entanglement among the emitters or any other partition of the system is absent \cite{Wolfe}. As can be seen, the LQU presents a double sudden change, the first one at time $t_{I}$ and the second one at $t_{F}$, with $\Delta_{DSC} = t_{F} -t_{I} > 0$. In the intervals $t_{I} \ge t \ge 0$ and $\infty > t \ge t_{F}$ the LQU is minimized by the local observable $\sigma_{z}^{l}$, while in the interval defined by $\Delta_{DSC}$ it is minimized by $\sigma_{x}^{l}$. Once the LQU behavior is superposed to that of $P_{corr}$ for earlier and later stages of the phenomenon (see the blue dotted line and red dashed line in FIG.1), we can conclude that in these time intervals the quantum correlations between each emitter and the remaining system is responsible for the quantum uncertainty on the observable $\sigma_{z}^{l}$, and consequently, responsible for the time behavior of the correlation radiation rate $P_{corr}(t)$. These correlations are predominant during the major part of the superradiance phenomenon for high values of $N$, provided that $\Delta t_{DSC} \propto 1/N$. Meanwhile, in the time window $\Delta t_{DSC}$ the quantum correlations between each emitter and the remaining system captured by LQU is weakened, allowing the emergence of other types of correlations among the emitters. 

\emph{WYSI for global observables} -- The WYSI for the global observables $J_{x}$, $J_{y}$, and $J_{z}$ can be obtained using Eqs. (\ref{eq:WI}) and (\ref{eq:Operators})-(\ref{eq:rho}). For the observable $J_{z}$ the WYSI has a trivial solution $I(\rho (t),J_{z} )=0$, since it commutes with the state of the system $\rho (t)$ at all times. This occurs because the state of the system is diagonal in the $J_{z}$ basis. Therefore, $\rho(t)$ is incoherent in the Dicke basis at any time \cite{Baumgratz}. Analysing the symmetry of the operators $J_{x}$ and  $J_{y}$ with relation to the state $\rho(t)$, it is easy to verify that $I(\rho(t),J_{x} )=I(\rho(t),J_{y})$, which allows the calculation of the WYSI only once. Thus, we obtain for the total polarization operator $J_{x}$ the expression
\begin{multline}
	I(\rho(t),J_{x} ) =\frac{1}{2} \{J(J+1) - \sum_{M=-J}^J \{ p_{J,M}(t)M^{2}\\
 - \sqrt{p_{J,M}(t)p_{J,M-1}(t)}\left[ J(J+1) - M(M-1)\right]   \} \},
\end{multline} 
where we have used the relations between $J_{x}$ and the ladder operators $J_{\pm}$, introduced above, and the well known relation $J_{\pm} |J,M\ket = \sqrt{J(J+1)-M(M\pm1)} |J,M\pm1\ket$.

In FIG. 2 we plot the normalized radiated power $P(t)/500$ and the normalized WYSI $4I(\rho(t),J_{x} )/50$ as functions of the dimensionless time $\gamma t$ for $N=50$ emitters. It is possible to observe that the total radiated power achieves its maximum value at time $t_{max}$, which is very close to the time $t_{min}$ in which the WYSI for $J_{x}$ achieves its minimum value. The time behavior of the normalized $P(t)$ and $I(\rho(t),J_{x} )$ is very similar for different values of $N$. However, as $N$ increases, the time $t_{min}$ becomes closer to $t_{max}$, so that in the classical limit ($N \rightarrow \infty$) they coincide, as shown in FIG. 3. Therefore, beyond the WYSI for the local observable $\sigma_{z}^{l}$, the WYSI for the operator $J_{x}$ can be used as a figure of merit to characterize the time of maximum emission of radiation in the Dicke Superradiance for large values of $N$.
\begin{figure}[h]
     \includegraphics[scale=0.43]{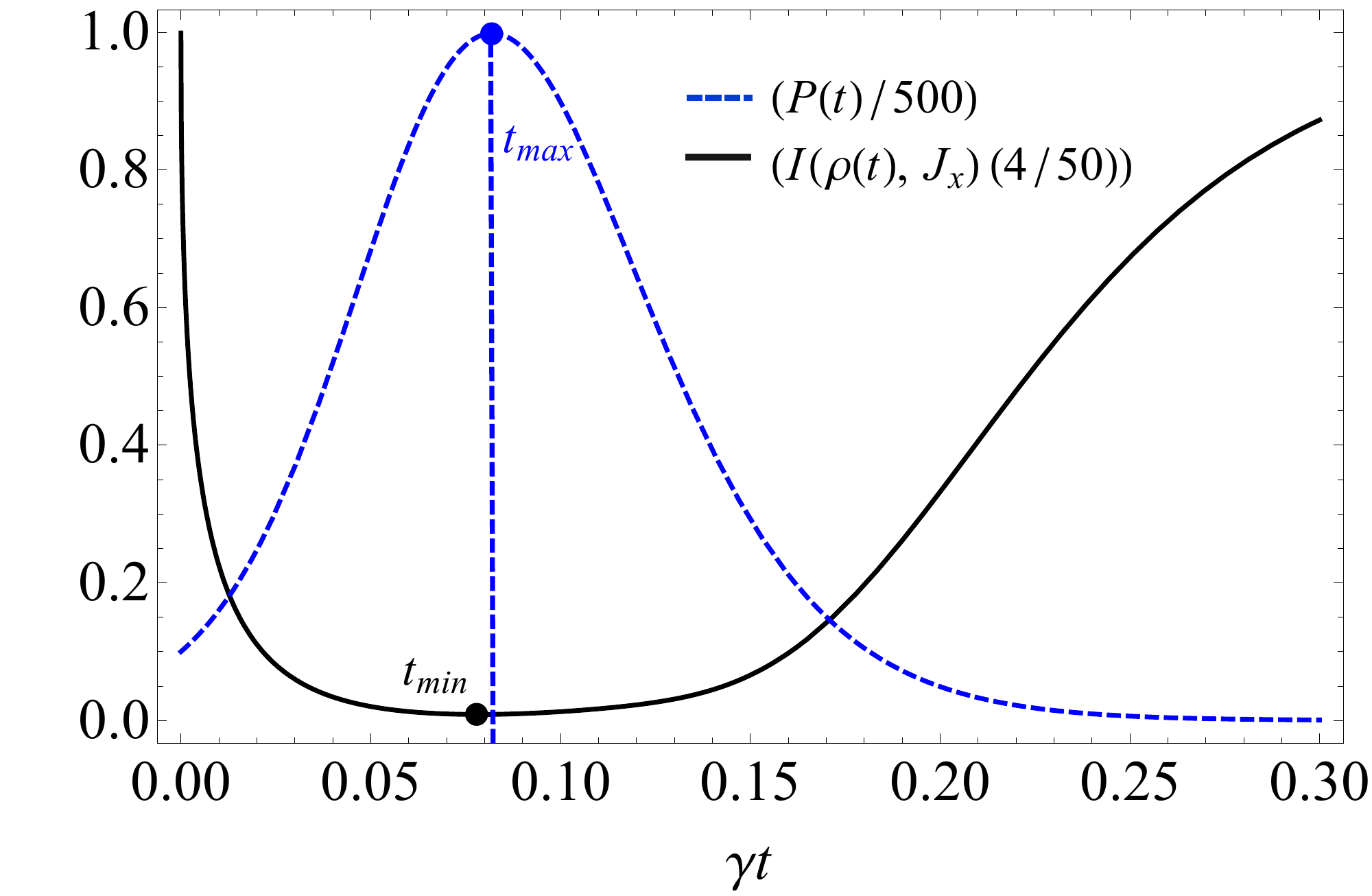}
     \label{WYJx}
 \caption{\small (Color online) Normalized radiated power $P(t)/500$ (dashed blue line) and the normalized WYSI $4I(\rho(t),J_{x} )/50$ (solid black line) as function of the dimensionless time $\gamma t$ for $N=50$ emitters. The vertical dashed blue line indicates the time of maximum emission of radiation $t_{max}$, while the solid black circle bellow shows the time in which the WYSI achieves its minimum value, $t_{min}$.}
\end{figure}

The previous result can immediately be generalized by noticing that $lim_{N \rightarrow \infty} t_{min} = t_{max}$ has its origin in the state of the system and is independent of the symmetric operator used to evaluate the WYSI. From FIG. 1 we observe that in the interval $\Delta t_{DSC}$ all Dicke states are almost equally occupied,  $p_{J,M} (t \backsim t_{max}) \backsimeq 1/(N+1)$. This approximation becomes better for large values of $N$, since the contribution of states near to $|J,J\ket$ and $|J,-J\ket$, which are dominant at the beginning and at the end of the process, is weakened. According to Eq. (\ref{eq:rho}), it follows that $\lim_{N \rightarrow \infty} \rho (t \backsim t_{max}) \simeq \textbf{1}_{sym}/(N+1)$, where $\textbf{1}_{sym}$ is the identity operator in the symmetric subspace. Therefore, for any symmetric global operator $K_{sym}$, i.e., an operator which takes a Dicke state into another Dicke state, $[\textbf{1}_{sym},K_{sym}]=0$, so that $\lim_{N \rightarrow \infty} I(\rho(t \backsim t_{max}),K_{sym}) = 0$. As the WYSI is a positive semi-definite real function, we conclude that at $t_{max}$ it is minimum. Of course that symmetric operators that commute with $\rho(t)$ at all times are not interesting, as in the case of $J_{z}$. Summarizing, the fact that the radiated power achieves its maximum value at the same time the WYSI for any symmetric global operator achieves its minimum value, suggests that the later can be used as a figure of merit to characterize the time of maximum emission of radiation in the Dicke Superradiance for $N \rightarrow \infty$. 

\begin{figure}[h]
     \includegraphics[scale=0.42]{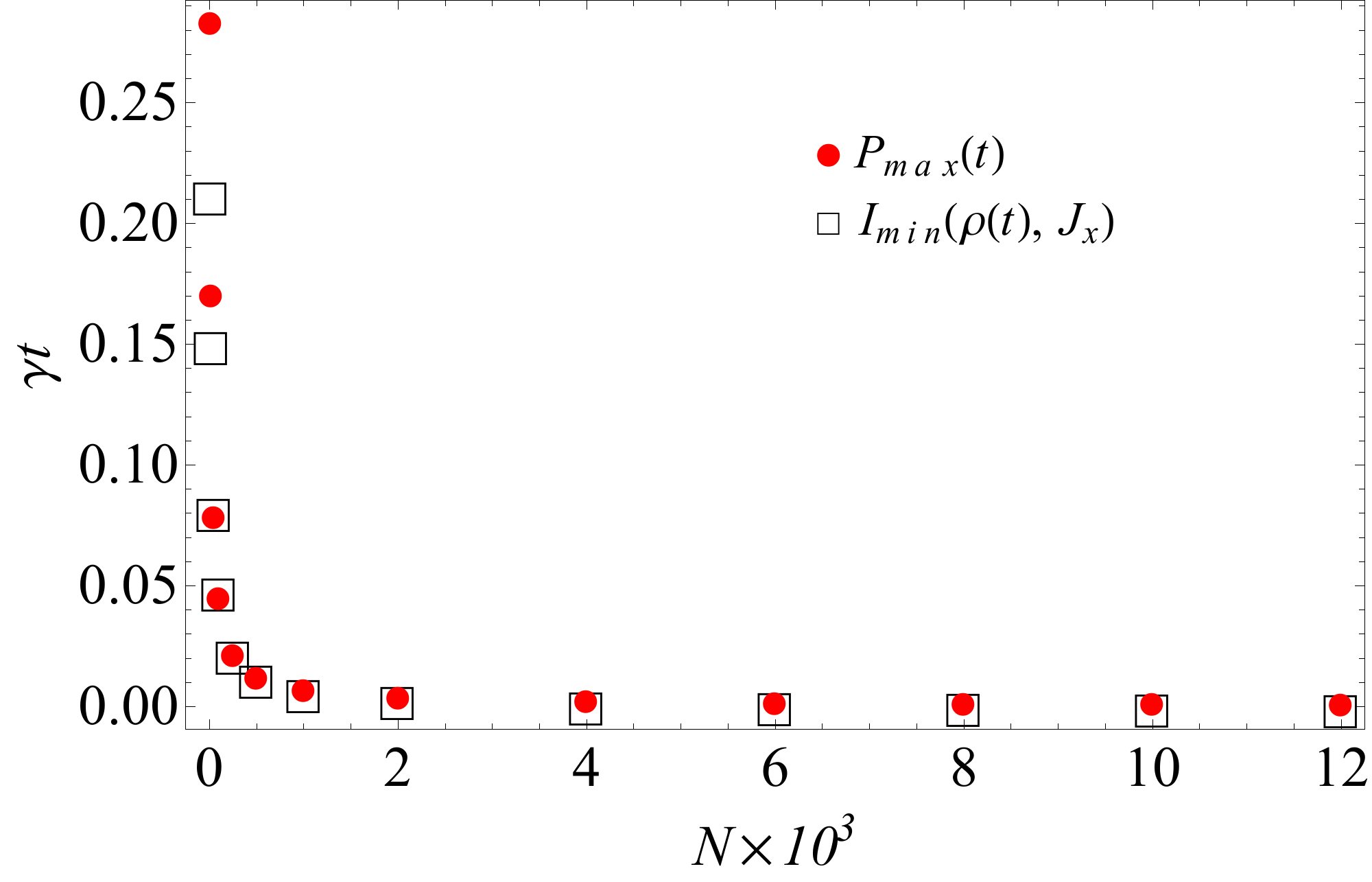}
     \label{scalability}
 \caption{\small (Color online) Dimensionless time $\gamma t$ in which the radiated power achieves its maximum value, $ P_{max}(t) $, and the time in which the WYSI for $J_{x}$ operator achieves its minimum value, $I_{min}(\rho(t),J_{x} )$, as function of the number $N$ of atoms in the sample.}
\end{figure}

\emph{Global versus local observables} -- The classical limit in the Dicke superradiance is taken over global observables for a large number of emitters ($N \rightarrow \infty$). Such limit has been analyzed from different points of views. In Refs. \cite{Gross,Haake} it is assumed that the classical limit appears when the fluctuations measured by the standard deviation $\sqrt{Var(\rho,K)}$ become very small compared to the value of the observable being analyzed. For instance, at the early stages of the evolution the number of emitted photons $\bra J - J_{z} \ket$ is very small, with values of the same order of magnitude of its standard deviation. This implies that the classical regime occurs after the early stages of the evolution, provided that the number of emitted photons always increase until $N$. We point out that measures as variance, also account for classical uncertainties, blurring quantum effects. As the WYSI can be used as a good indicator of the existence of quantum features in the system, since it captures part or the whole quantum uncertainty of a given observable, we will use it to analyze the quantum-classical crossover. As shown above, the WYSI for any symmetric global operator $K_{sym}$ in the interval $\Delta t_{DSC}$ is null for a large number of emitters. This means that all symmetric global observables behave as classical random variables, even though the emitters are strongly quantum correlated during the time interval $\Delta t_{DSC}$. As far as we know, this result on quantum-classical crossover is quite intriguing, since we are not finding conditions where the whole quantum system behaves classically \cite{Gross,Haake}, or just part of it \cite{Rossatto}, but we found a situation where the behavior of the system is determined by the scale we look at it, i.e., to the macro-world or micro-world. 

\emph{Conclusions and perspectives} -- We have shown that the quantum uncertainty of the local observable related to the population of each emitter is responsible for the time behavior of the Dicke superradiance. The quantum correlations measured by LQU, which is a discord-like quantum correlation, present a double sudden change induced by environment. During the time window between these two sudden changes the symmetric global observables of the system behave classically in the limit $N \rightarrow \infty$, although the emitters are strongly quantum correlated. It would be interesting to analyze the quantum correlations in generalized models of superradiance \cite{Tang}.

The authors thank to M. S. Sarandy for reading the manuscript and for helpful comments. The financial support for this work was provided by the Brazilian funding agencies CNPq, CAPES, and also by the Brazilian National Institute of Science and Technology for Quantum Information.

\bibliographystyle{unsrt} 

\begin{thebibliography}{99}

\bibitem{Shor}{P. W. Shor, \emph{SIAM J. Comput.} \textbf{26}, 1484 (1997).}

\bibitem{Feynman}{R. P. Feynman, \emph{Inter. J. Theo. Phys.} \textbf{21}, 467 (1982).}

\bibitem{Hallgren}{S. Hallgren, \emph{Proc. of the 34th Symp. on the Theory of Computing} \textbf{653} (2002).}

\bibitem{Freedman}{M. H. Freedman, A. Kitaev, and Z. Wang, \emph{Comm.Math. Phys.} \textbf{227}, 587 (2002).}

\bibitem{Aharonov}{D. Aharonov, V. Jones, and S. Landau, \emph{Proc. 38th Annual ACMSymp. on Theory of computing} \textbf{427} (2006).}

\bibitem{Harrow}{A. W. Harrow, A. Hassidim, and S. Lloyd, \emph{Phys. Rev. Lett.} \textbf{103}, 150502 (2009).}

\bibitem{Jozsa}{R. Jozsa and N. Linden, \emph{Proc. R. Soc. A} \textbf{459}, 2011 (2003).}

\bibitem{Ollivier}{H. Ollivier and W. H. Zurek, \emph{Phys. Rev. Lett.} \textbf{88}, 017901 (2001).}
   
 \bibitem{Henderson}{L. Henderson and V. Vedral, \emph{J. Phys. A} \textbf{34}, 6899 (2001).}  
 
\bibitem{Datta}{A. Datta, A. Shaji, and C. M. Caves, \emph{Phys. Rev. Lett.} \textbf{100}, 050502 (2008).}

 \bibitem{Zurek}{W. H. Zurek, \emph{Phys. Rev. A} \textbf{67}, 012320 (2003).}
 
 \bibitem{Piani2011}{M. Piani, et al., \emph{Phys. Rev. Lett.} \textbf{106}, 220403 (2011).}
 
 \bibitem{Modi2011}{K. Modi, H. Cable, M. Williamson, and V. Vedral, \emph{Phys. Rev. X} \textbf{1} 021022 (2011).}
 
 \bibitem{Meznaric}{S. Meznaric, S. R. Clark, and A. Datta, \emph{Phys. Rev. Lett.} \textbf{110}, 070502 (2013).}
 
 \bibitem{Madhok}{V. Madhok and A. Datta, \emph{Phys. Rev. A} \textbf{83} 032323 (2011).}
 
 \bibitem{Piani2014}{M. Piani, V. Narasimhachar, and J. Calsamiglia, \emph{New J. Phys.} \textbf{16}, 113001 (2014).}

\bibitem{Modi}{K. Modi, et al., \emph{Rev. Mod. Phys.} \textbf{84}, 1655 (2012). See also the references therein.}
   
\bibitem{Celeri} L. C. Celeri, J. Maziero, and R. M. Serra,\emph{Int. J. Quant. Inform.} \textbf{9}, 1837 (2011).
   
\bibitem{Dicke}{R. H. Dicke, \emph{Physical Review} \textbf{93}, 99 (1953).}

\bibitem{Bohnet}{J. G. Bohnet, et al.,  \emph{Nature} \textbf{484}, 78 (2012).}

\bibitem{Monshouwer}{R. Monshouwer, M. Abrahamsson, F. van Mourik, and R. J. van Grondelle, \emph{J. Phys. Chem. B} \textbf{101}, 7241 (1997).}

\bibitem{Zhao}{Y. Zhao, et al., \emph{J. Phys. Chem. B} \textbf{103}, 3954 (1999)}.

\bibitem{Celardo}{G. L. Celardo, G. G. Giusteri, and F. Borgonovi, \emph{Phys. Rev. B} \textbf{90}, 075113 (2014).}

\bibitem{Wolfe}E. Wolfe and S. F. Yelin, \emph{Phys. Rev. Lett.} \textbf{112}, 140402 (2014).
   
\bibitem{Wignera}{E. P. Wigner, \emph{Z. Phys.} \textbf{133}, 101 (1952).}
      
\bibitem{Araki}{H. Araki and M. M.  Yanase, \emph{Phys. Rev.} \textbf{120}, 622 (1960).}
      
\bibitem{Wignerb}{E. P. Wigner and M. M. Yanase, \emph{Proc. Natl. Acad. Sci. USA} \textbf{49}, 910 (1963).}
 
\bibitem{OBS}{Some care must be taken in order to increase the number of emitters in the sample ($N \rightarrow \infty$), since the idealized Dicke model of superradiance remains valid for samples whose distances between emitters are shorter than the emission wavelength.} 
     
\bibitem{Girolamia}{D. Girolami, T. Tufarelli, and G. Adesso \emph{Phys. Rev. Lett.} \textbf{110}, 240402 (2013).}

\bibitem{Gross}{M. Gross and S. Haroche, \emph{Phys. Rep.} \textbf{93}, 301 (1982).}
     
\bibitem{Agarwal}{G. S. Agarwal, \emph{Quantum statistical theories of spontaneous emission and their relation to other approaches}, Springer Tracts in Modern Physics, Vol. 70., ed. G. H\"{o}hler, Springer-Verlag, Berlin Heidelberg, 1974.}
      
\bibitem{Delanty}{M. Delanty, S. Rebic, J. Twamley, arXiv:1107.5080v1, (2011).}
       
\bibitem{Fanchini}{F. F. Fanchini, T. Werlang, C. A. Brasil, L. G. E. Arruda, A. O. Caldeira, \emph{Phys. Rev. A} \textbf{81}, 052107 (2010).}  
    
\bibitem{Montealegre}{J. D. Montealegre, F. M. Paula, A. Saguia, and M. S. Sarandy, \emph{Phys. Rev. A} \textbf{87}, 042115 (2013).}
                
\bibitem{Paula}{F. M. Paula, et al., \emph{Phys. Rev. Lett.} \textbf{111}, 250401 (2013).}
       
\bibitem{Baumgratz}{T. Baumgratz, M. Cramer, and M. B. Plenio, \emph{Phys. Rev. Lett.} \textbf{113} 140401, (2014).}
      
\bibitem{Haake}{F. Haake and R. J. Glauber, \emph{Phys. Rev. A} \textbf{5}, 1457 (1972).}    

\bibitem{Rossatto}{D. Z. Rossatto, T. Werlang, E. I. Duzzioni, and C. J. Villas-Boas, \emph{Phys. Rev. Lett.} \textbf{107} 153601, (2011).}  

\bibitem{Tang}{S.-Q. Tang, J.-B. Yuan, L.-M. Kuang, and X.-W. Wang, \emph{Quantum Inf. Process.} \textbf{14} 2883 (2015).}

       


\end{thebibliography}
\addcontentsline{toc}{chapter}{\protect\numberline{}Bibliografia}

\end{document}